\documentclass[12pt]{article}
\begin{document}
\newcommand{\br}{{\bf r}}
\newcommand{\grad}{\mbox{\boldmath$\nabla$}}
\newcommand{\bdiv}{\mbox{\boldmath$\nabla\cdot$}}
\newcommand{\curl}{\mbox{\boldmath$\nabla\times$}}
\newcommand{\bcdot}{\mbox{\boldmath$\cdot$}}
\newcommand{\btimes}{\mbox{\boldmath$\times$}}
\newcommand{\btau}{\mbox{\boldmath$\tau$}}
\newcommand{\btheta}{\mbox{\boldmath$\theta$}}
\newcommand{\bmu}{\mbox{\boldmath$\mu$}}
\newcommand{\bepsilon}{\mbox{\boldmath$\epsilon$}}
\newcommand{\bcj}{\mbox{\boldmath$\cal J$}}
\newcommand{\bcf}{\mbox{\boldmath$\cal F$}}
\newcommand{\bbeta}{\mbox{\boldmath$\beta$}}
\newcommand{\lbar}{\lambda\hspace{-.09in}^-}
\newcommand{\bcp}{\mbox{\boldmath$\cal P$}}
\newcommand{\bco}{\mbox{\boldmath$\omega$}}
\newcommand{\brho}{\mbox{\boldmath$\rho$}}
\newcommand{\balpha}{\mbox{\boldmath$\alpha$}}
\title{Discrete classical electromagnetism}
\author{Jerrold Franklin\footnote{Internet address:
Jerry.F@TEMPLE.EDU}\\
Department of Physics\\
Temple University, Philadelphia, PA 19122-6082}
\date{\today}
\maketitle
\begin{abstract}
The electric charges that make up charge distributions and currents in classical electromagnetism (EM) are point charge electrons with a fixed finite charge.  This means that a truly continuous charge distribution or current cannot be formed because that would require infinitesimal charges. If the number of electrons in any distribution is very large, the distributions can be approximately continuous, and can be used in many applications, but the discrete character of the distributions has a major effect on other physical effects.   
We consider two of those effects in this paper.  Does a point charge have infinite self-energy, and does a moving current loop have an electric dipole moment?
\end{abstract}
\section{Introduction}

Classical electromagnetism studies the electromagntic interactions of charged particles and, presumably, continuous charge distributions. But what are `continuous' charge distributions made of?
Is there a charged soup without charged particles or are continuous charge distributions made up of discrete charged particles of infintesimal size? 

In this paper, we consider that there is another alernative.  Electric charges and currents are made up of electrons that are point particles with a fixed, finite charge.  In this paper, we examine the consequences of classical electromagnetism (EM) without truly continuous charge distributions.

\section{Electromagnetic energy of a  charge distribution.},

A collection of N individual point charges has Coulomb energy\footnote{We define the Coulomb energy as the energy required to bring a point charge from infinite distance to its present position with a negligible velocity.}\footnote{This equation is in many EM textbooks, for instance, Eq,~(2.41) of \cite{dg}.}\footnote{We are using Gaussian units with c=1.}
\begin{equation}
U=\frac{1}{2}\sum_i^N\sum_{j\ne i}^N\frac{q_iq_j}{|{\bf r}_i-{\bf r}_j|}. 
\end{equation}
The condition, $j\ne i$, is necessary because the first charge put in place has no Coulomb energy.
This equation is a discrete sum of $N$ point charges.
One distinctive feature is that all of the interactions are two body interactions. There is no self-energy of any kind.

In order to consider more general cases,
 most sources have tried
to introduce
a continuous charge distribution.
The standard way to convert the discrete sum in Eq.~(1) into a continuous distribution is to take the limits,
\begin{equation}
 N\rightarrow\infty,\quad q_i\rightarrow 0,
\end{equation}
Also, each point charge should be infinitesimally close to all adjacent charges.

However, charge and current distributions in classical electromagnetism are composed of electrons.
There just are no other particles available.                                           
Even composite particles in nuclear or weak interactions have charges that are integral multiples of the electron charge.  An important point for this paper is that there are no actual particles with infinitesimal charge.
Even if there were infinitesimal charges comprising a continuous distribution, point charges 
(like electrons) would still be a discrete contribution to the distribution.  
 
We take electrons to be point particles\footnote{There have been attempts, notably by Lorentz\cite{hl}, to consider electrons as charged particles with finite or infinitesimal radius. But these models have failed for a number of well understood reasons.} of fixed charge, $e$, that cannot be reduced or considered infinitesimal. 
 That means that there is no truly continuous charge distribution in classical electromagnetism.
 
 The distributions,  $\rho({\bf r})$ and ${\bf j(r)}$,\footnote{We have shown in \cite{j} that an electric current density is actually composed of moving electrons, and is not continuous.} 
  in Maxwell's equations, are also not truly continuous distributions.
but, can often be approximated as continuous. 
Making $N$ very large (of the order of Avogradro's constant), seems to work in many cases.
However, for some important cases, the fact that the charge and current densities are actually discrete, produces sharp differences in physical results which we consider below. .

\section{Consequences of absence of continuous distributions.}

In earlier papers\cite{j,j2}, we have described some consequences of the discrete nature of charge and current distributions:\\

\subsection{Self energy in charge distributions?}
The notion of an infinite electromagnetic (EM) self-energy of point charges (presumably electrons) is often assumed. See, for instance,\cite{dg,rf,jdj}. However, each of these sources acknowledge that they don't understand that result.
For instance, Griffiths writes, ``The infinite energy of a point charge is a recurring source of embarrassment
for electromagnetic theory".\footnote{\cite{dg} page 96.},
and Feynman states,
``We must conclude that the idea of locating the energy in the field is inconsistent with the assumption of the existence of point charges.",\footnote{\cite{rf} Section 8-6.}

How can they be so dismissive of what they claim to have derived?
 In this Section, we show that, in fact, electrons must be point particles with no EM self-energy.  

The textbook derivations of an infinite self-energy of point charges
are generally based on a continuous charge distribution, They start by writing Eq.~(1) for a single point charge at a position, $\bf r$, as\footnote{This is Eq.~(2.43) of \cite{dg}.}
\begin{equation}
U_i=q_i\phi({\bf r}).
\end{equation} 

Did you notice that this introduction of the potential, $\phi$, has lost the important instruction, $j\ne i$, in  Eq.~(1), from which it was was derived? In fact, Eq.~(3) is contradicted by the example discussed in Section (2.42) of \cite{dg}, which has no $q_iq_i$, self-energy term. Self-energy has spuriously been put into the EM energy.

It is clear from the above observation that 
 Eq.~(3) should be given as
\begin{equation}
U_i=q_i\tilde{\phi}({\bf r}),
\end{equation} where $\tilde{\phi}$ is a potential that does not include a contribution due to the point charge itself, since this would have put a term $q_iq_i$ into Eq.~(1). 

Equation (3) is usually used to modify discrete sums to integrals over `continuous' charge distributions.\footnote{Actually, integrals are usually evaluated by discrete sums.}        
For instance, 
in Section 1.11, of Jackson\cite{jdj},
his discrete equation (1.51),\footnote{This is just our Eq.~(1)}.
\begin{equation}
U=\frac{1}{2}\sum_i^N\sum_j^N\frac{q_iq_j}{|{\bf r}_i-{\bf r}_j|}, 
\end{equation}
is given with the understanding ``that  i = j terms (infinite `self-energy' terms) are omitted in the
double sum." 
But, he replaces this discrete sum with the integral,
\begin{equation}
U=\frac{1}{2}\int\int\frac{\rho({\bf r})\rho({\bf r'})d^3r d^3 r'}{|{\bf r'}-{\bf r}|},
\end{equation}
losing the fact that the same point charge cannot be in each charge density.               

EM textbooks also try to introduce self-energy by finding the energy in the electric field of the point charge.
For instance, Griffiths goes from his Eq.~(2.41) by several steps, leading to his Eq.~(2.45) for the energy as
\begin{equation}
U=\frac{1}{8\pi}\int E^2 d^3r.
\end {equation}
This would give infinite self-energy to a point charge, again mistakenly.
The energy in terms of the electric field should be given as
\begin{equation}
U=\frac{1}{8\pi}\int{\bf \tilde{E}\bcdot\tilde{E}} d^3r,
\end {equation}
ruling out-self energy because any term like $q_iq_i$ would not be permitted.

\subsection{Induced electric dipole moment for a moving current loop?}

It is well known that a complete Lorentz transformation of a
point charge is a two step operation.
For a single point charge, EM textbooks first transform  
  Coulomb's law for the electric field of the charge at rest in a system S$'$
to a system, S, where the charge is moving with velocity, $V$.
Then, in a second step, the position variable, $\bf r'$, has to be Lorentz transformed
to $\bf r$ in the moving system.

It has not generally been realized that 
Lorentz transforming a current distribution must also be done in two steps.
First, the current distribution must be transformed, and then the individual point charges in the current distribution must be transformed.

The first step is to transform
the charge-current density four-vector,
 $[\rho'({\bf r'},t'),{\bf j'(r'},t')]$, 
with a Lorentz transformation from a system S$'$ in which a neutral current loop, 
(with $\rho'=0$), is at rest,
to a system S in which the current loop is moving with a velocity $\bf V$. The charge and current densities will transform to
\begin{eqnarray}
\rho({\bf r'},t')&=&\gamma{\bf V\cdot j'(r',}t')\label{rp}\\
{\bf j(r',}t')&=&\gamma{\bf j'(r',}t').
\end{eqnarray}

At this point most derivations stop,\footnote{See, for instance\cite{mm}.}
apparently considering $\rho({\bf r'},t')$ to be the final continuous charge distribution. This suggests
 the incorrect conclusion that a moving current loop has an electric charge distribution and electric dipole moment.
This would result in an electric force and torque on the moving loop.
This force and torque on an originally neutral loop has led to what has been called the Mansuripur paradox.\cite{mm}

The necessary second step of Lorentz transforming the discrete, moving, electrons of the current, results in the complete, two step, Lorentz transformation result.  
In the second step, the electrons take a finite time to go from one end of a sampling cell to the other end.
This means the electrons are counted at different times in the original rest system of the loop.
 This turns out to remove any change in the charge density, which remains at zero.\footnote{The somewhat complicated derivation of this is shown in detail in \cite{j}.}
 Thus, the moving current loop has no charge density and no electric dipole moment.  There is no Mansuripur paradox.

\section{Conclusion}

There are no truly continuous charge or current distributions in classical electromagnetism.
This is because the only charge carriers are electrons with fixed charges that cannot be infinitesimal, 
as is required for taking a continuous limit.
Consequently, we have, shown that there is no self-energy (infinite or not) of point charges, and a moving current loop does not have an electric dipole moment.

Readers are encouraged to suggest other spurious results caused by ignoring the discrete nature of charge and current distributions in classical electromagnetism.


\begin{thebibliography}{9}
 \bibitem{dg}D. J. Griffiths   {\it Introduction to Electrodynamics,
   4th Edn.} (Addison Wesley) 
  \bibitem{rf}R. Feynman, R. B. Leighton, M. Sands {\it The Feynman Lectures on Physics},
 vol. 2. (Addison-Wesley). 
\bibitem{jdj}J. D. Jackson {\it Classical Electrodynamics 3rd Edn,} 
(John Wiley \& Sons)  
 \bibitem{hl}H. A. Lorentz, ``Versuch einer Theorie der electrischen und optischen Erscheinungen in bewegten Körpern  [Attempt of a Theory of Electrical and Optical Phenomena in Moving Bodies]", Leiden: E.J. Brill
\bibitem{j2}J. Franklin, ``Infinite self-energy?"
DOI: 10.48550/ARXIV.2511.19571
\bibitem{j}J. Franklin, ``Complete Lorentz transformation of a charge-current density",
International Journal of Modern Physics A
2020-04-30 .\\
DOI: 10.1142/S0217751X2050061X
\bibitem{pp}W. K. H. Panofsky and M. Phillips {\it Classical Electricity and Magnetism}, 2nd Edition (Addison-Wesley, AW.).
\bibitem{mm}M. Mansuripur,  ``M. Mansuripur, “Trouble with the Lorentz Law of Force: Incompatibility
with special relativity and momentum conservation",
Phys. Rev. Lett. 108, 193901 (2012).
\end{thebibliography}
\end{document}